# Two New Variations on the Twin Pseudoparadox

Jean-Marc Lévy-Leblond
University of Nice-Sophia Antipolis

**Abstract**. Two new scenarios are proposed which generalize the standard story leading to the pseudoparadox of the Einsteinian relativistic twins, thereby enabling some deeper understanding.
First, Aesop's fable "The Hare and the Tortoise" is considered in the light of Einsteinian chronogeometry. It is then shown that the Hare, while arriving later than the Tortoise, may still be the winner of the race — or at least may consider itself to be.
Second, the situation is considered where the twin initially left at home decides to catch up his brother during his travel. Can they meet so that they may celebrate a common anniversary and recover the same age?

## Is it worthwhile (re)$^n$visiting the Twin Paradox?

No question perhaps in modern physics has been discussed as much as the (in)famous twin paradox in Einsteinian relativity. Since it was first mentioned by Einstein and others forefathers[1], it has been the subject of hundreds of papers and continues to this day to supply a continuous flurry of articles[2]. While some recent papers are concerned with generalizations to general relativistic and/or quantum situations, by far most of the literature on the subject, when written from a pedagogical or philosophical point of view, concentrates on the standard situation where one of the twins makes a round trip from Earth to some far away star while the other stays at home. It may be argued that this is too simple a situation for enabling the ingenuous layperson or beginner to grasp the full depth and necessity of the reason for the differential aging of the twins upon their meeting. Rather than proposing a new conceptual discussion, we will sketch here two different scenarios exhibiting the phenomenon in a more general context.

## The Hare, the Tortoise and Einstein

*"The Hare laughed at the Tortoise's feet but the Tortoise declared, 'I will beat you in a race!' The Hare replied, 'Those are just words. Race with me, and you'll see! Who will mark out the track and serve as our umpire?' 'The Fox,' replied the Tortoise, 'since she is honest and highly intelligent.' When the time for the race had been decided upon, the Tortoise did not delay, but immediately took off down the race course. The Hare, however, lay down to take a nap, confident in the speed of his feet. Then, when the Hare eventually made his way to the finish line, he found that the Tortoise had already won." Aesop's Fables[3]*

It is clear that this story happened more than two thousand years ago. For nowadays, the Hare would know better!

Let us take a closer look at the scene taking place on the finishing line. Suppose that the Fox, the Hare and the Tortoise each carry chronometers which have been carefully synchronized before the start of the race. The Fox indeed sees the Tortoise crossing the finish line first, and tells him that his running time has been 20 minutes.

---

[*] jmll@unice.fr

The Tortoise takes a look at his own watch, and, somewhat puzzled, replies that he has been running for 16 minutes only. Before they can argue over the discrepancy and settle the matter, arrives the Hare. "One minute too late!" announces the Fox. "Not at all, rejoins the Hare triumphantly. I have beaten the Tortoise by one minute! It has taken me but 15 minutes to get here from the starting signal, although I enjoyed a small nap of 6 minutes before moving on. Don't you know your Einstein?"

Now for a few easy calculations. Let us denote by $d$ the length of the race, $v$ and $v'$ respectively the speeds of the Tortoise and the Hare, with the obvious inequality $v' > v$. Suppose the Hare only starts running after a nap delay $\delta$ with respect to the starting signal, while the Tortoise runs off immediately. It is clear that the time elapsed between the start of the race and the arrival at the finish line, as judged by the Fox, are:

$t = d/v$

$t' = \delta + d/v'$

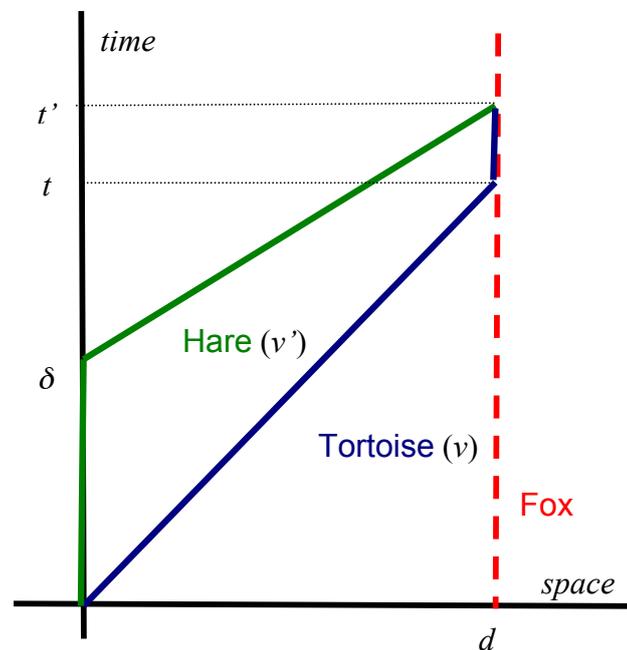

The arrivals of the Tortoise and the Hare will be separated by a time interval

$$\Delta t = t' - t = \delta + \left(\frac{1}{v'} - \frac{1}{v}\right)d,$$

meaning that the Hare will arrive after the Tortoise ($\Delta t > 0$) if he dozes longer than a critical delay

$$\delta_c = \left(\frac{1}{v} - \frac{1}{v'}\right)d.$$

But let us consider now the race from the point of view of the runners, taking into account that they measure their *proper* times.

Elementary Einsteinian relativity tells us that the Tortoise will register his running time as

$$\tau = d\frac{\sqrt{1-v^2}}{v}$$

while the Hare's watch will show upon arrival the elapsed time

$$\tau' = \delta + d\frac{\sqrt{1-v'^2}}{v'}$$

(we have chosen units such that the limit speed is unity: $c = 1$).
The time difference between these durations is

$$\Delta\tau = \tau' - \tau = \delta + (\frac{\sqrt{1-v'^2}}{v'} - \frac{\sqrt{1-v^2}}{v})d,$$

so that the Hare's proper running time will be longer than the Tortoise's ($\Delta\tau > 0$) if his nap is longer than a new critical (Einsteinian relativistic) delay

$$\delta_r = \left(\frac{\sqrt{1-v^2}}{v} - \frac{\sqrt{1-v'^2}}{v'}\right)d.$$

The point now is that one always have
$$\delta_r > \delta_c,$$
so that, if the Hare starts running between times $\delta_c$ and $\delta_r$, he will arrive after the Tortoise, and nevertheless he will have spent a shorter (proper) time between the start and the arrival of the race.
The simplest way to prove that $\delta_r > \delta_c$ is to switch from the velocities $v$ and $v'$ to the rapidities $\varphi = \tanh^{-1} v$ and $\varphi' = \tanh^{-1} v'$. With these notations, the Hare's critical delays respectively become:

$$\delta_c = \left(\frac{1}{\tanh\varphi} - \frac{1}{\tanh\varphi'}\right)d$$

and

$$\delta_r = \left(\frac{1}{\sinh\varphi} - \frac{1}{\sinh\varphi'}\right)d.$$

But then

$$\delta_r - \delta_c = \left(\frac{1}{\sinh\varphi} - \frac{1}{\sinh\varphi'}\right)d - \left(\frac{1}{\tanh\varphi} - \frac{1}{\tanh\varphi'}\right)d$$
$$= \left(\frac{1-\cosh\varphi}{\sinh\varphi} - \frac{1-\cosh\varphi'}{\sinh\varphi'}\right)d = \left[\tanh(\varphi'/2) - \tanh(\varphi/2)\right]d$$

which shows that
$$\delta_r - \delta_c > 0, \quad \text{since} \quad \varphi' > \varphi.$$

In other words, it is always possible for the Hare, thanks to Einstein, to take a longer nap (up to $\delta_r$) than Galileo and Newton would allow (not more than $\delta_c < \delta_r$). If his starting time is such that $\delta_c < \delta < \delta_r$, the Hare's total napping and running (proper) time is shorter than the Tortoise's despite his later arrival. Note further that, since the Hare may evaluate the speed of the Tortoise when he observes his adversary running off, he is able to choose his own departure time so as to play this bad joke.
In order to obtain the numerical values quoted above, one has to take $v = 3/5$ and $v' = 4/5$ …and $d = 12$ light – minutes (that is, roughly 216 million kilometres, slightly more than the distance of the Earth to the Sun, which would require very "fast and furious" runners indeed).

**Can the relativistic twins still celebrate a common anniversary?**

*The starship* Argo *is to depart for the far away planet Kolchis with, among its crew, the twin brothers Cass-Tor and Paul-Lux as navigation officers. Unfortunately, at the last moment, Cass-Tor falls victim to a strange disease and cannot board. Left alone on Earth, he laments about his fate, all the more so since he knows that on returning, Paul-Lux will be much younger than him, thus breaking the symmetry between their lifespans. But some time after the departure of the expedition, the engineers discover that a flaw in the design of the ship threatens its safe return. It is then decided to launch an urgent rescue expedition. A second starship, named* Urgo, *is to join the* Argo *on its way and help its crew. Cass-Tor naturally enough asks for being part of this second expedition and is chosen to lead it. He then decides to plan the trip so as to meet Paul-Lux at a time when they will have reached the same age and can celebrate a common anniversary.*

Is this scenario possible at all?

Let us suppose that the starship *Argo* departs from Earth at time $t = 0$ with constant velocity $v$, reaches Kolchis situated at a distance $d$ away, and comes back with the very same velocity. We will neglect the acceleration and deceleration phases as well as the time spent around Kolchis. The *Argo* then is due to return at Earth time $T_E = 2d/v$, which would be the aging time of Cass-Tor, if he were to stay stranded on Earth. But the proper time of the trip for the *Argo*, that is the aging of Paul-Lux, will be $T_A = \gamma^{-1} T_E$, where $\gamma = (1 - v^2/c^2)^{-1/2} > 1$, so that $T_A < T_E$, which is but the standard "twin effect".

Now the starship *Urgo*, with Cass-Tor on board, leaves Earth at time $t_0$, with a velocity $v'$. The question is: can one choose the departure time $t_0$ and the velocity $v'$ so that, when the two ships meet, the proper times elapsed for Cass-Tor and Paul-Lux respectively since the departure of the *Argo* would be equal?

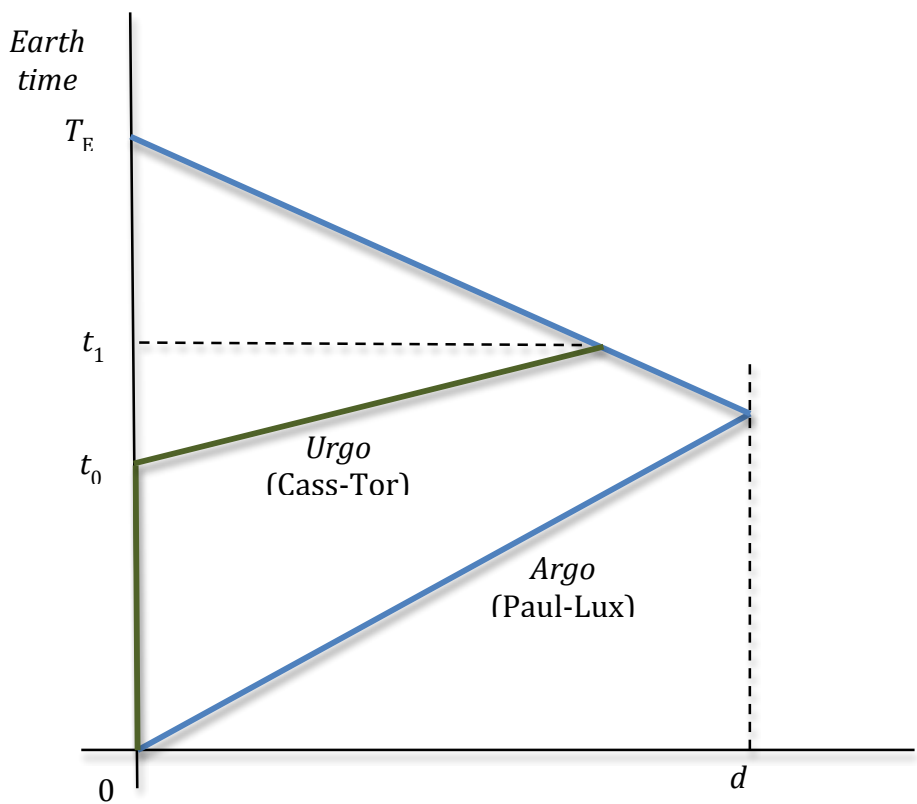

A first condition is obvious, before any calculation, that is, *Urgo*'s velocity must be higher than *Argo*'s: $v' > v$. Otherwise the shortening time coefficient $\gamma'^{-1}$ of Cass-Tor would be lower than the one of Paul-Lux $\gamma^{-1}$, preventing Cass-Tor from compensating part of the proper time lapse $t_0$ he spent on Earth. A second condition is, *Urgo* must meet *Argo* on its way back. Indeed, if *Urgo* catches up *Argo* before its turning point, Paul-Lux will have travelled a straight worldline, which would not be the case of Cass-Tor. The (proper) time spent on board of the *Argo* by Paul-Lux will then be longer than the time spent by Cass-Tor, consisting of the idle time spent on Earth plus the travel time of the *Urgo*. The two vessels thus have to meet at a time $t_1$, as measured on Earth, such that $t_1 > T_E / 2$.

If such is the case, the two ships will meet when the distance covered by *Urgo* from the Earth till the junction is equal to the distance remaining to be covered by *Argo* on its return trip, that is
$$v'(t_1 - t_0) = v(T_E - t_1).$$
The respective proper time intervals of Paul-Lux and Cass-Tor from their separation until their reunion are:
$$\tau_P = \gamma^{-1} t_1$$
$$\tau_C = t_0 + \gamma'^{-1}(t_1 - t_0).$$
For the twins to celebrate their common anniversary (with equal ages) upon meeting, they must have lived the same length of time, that is $\tau_C = \tau_P$. Denoting by $\tau$ this now common span of time, we may write
$$t_0 = \frac{\gamma' - \gamma}{\gamma' - 1} \tau$$
$$t_1 = \gamma \tau$$
hence
$$t_1 - t_0 = \gamma' \frac{\gamma - 1}{\gamma' - 1} \tau.$$
The meeting condition (1) can then be rewritten in terms of $\tau$, relating it to the total travel time $T_E$:
$$\tau = \left[\frac{v'}{v} \gamma' \frac{\gamma - 1}{\gamma' - 1} + \gamma\right]^{-1} T_E.$$
It turns out that, as is often the case, this rather awkward equation simplifies considerably when expressed with the help of the rapidities $\varphi = \tanh^{-1} v$ and $\varphi' = \tanh^{-1} v'$. Setting furthermore $s = \tanh(\varphi/2)$ and $s' = \tanh(\varphi'/2)$, we finally get the unexpectedly simple expressions
$$\tau = \frac{s'}{s + s'} \gamma^{-1} T_E = \frac{s'}{s + s'} T_A$$
The departure and meeting times $t_0$ and $t_1$ then are easily calculated:
$$t_0 = \frac{s' - s}{s'(1 + s^2)} T_E$$
$$t_1 = \frac{s'}{s + s'} T_E$$

Since $s' > s$, one checks that indeed $t_1 > T_E / 2$ or, equivalently, $\tau > T_A / 2$. More interesting is the fact that, due to the constraint $v' < c$, that is $s < 1$, there is a maximum delay time

$$t_{0\max} = \frac{1-s}{1+s^2} T_E$$

for *Urgo* to be launched after the departure of *Argo*, corresponding to the highest possible common aging for the twins to have a common anniversary

$$\tau_{\max} = \frac{1-s}{1+s^2} T_E \, .$$

The equality of $t_{0\max}$ and $\tau_{\max}$ is explained by the fact that when $v' \to c$, the proper time of Cass-Tor's journey goes to zero.

**Conclusion**

Of course there is really no more content in this exercise than in the classical Twin Paradox, but its scenario may offer some additional fun and interest. In any case, it offers a nice illustration that the difference of the time lapses between two spacetime events is much more general than the differential aging between two twins, one staying at home and the other travelling and returning. For in this conventional presentation, the reason for the age difference is often marred by the very special choice of the twins stories, one of them resting and the other one making a to and fro travel, with the tedious discussions going on about the effects of the acceleration, etc. A less dissymmetrical choice of the respective trajectories thus helps clearing up the situation. The matter is simply that the proper time between two events depends on the worldline followed from one to the other — exactly as, in ordinary space geometry, the distance between two points depends on the line connecting them. Einsteinian relativity is but a chronogeometry.

***

It is a pleasure to thank Elie During and Alexis Saint-Ours for organizing a colloquium on the Twin Paradox which furnished the incentive for the present exercises and for stimulating discussions, Françoise Balibar for her helpful comments, and Dominique Lacaze for correcting a shameful (but fortunately unconsequential) error.

---

[1] Albert Einstein, *Ann. der Physik.*, vol. XVII, 1905, pp. 891-921 ; *Naturforschende Gesellschaft*, n°56, 1911, pp 1–14.
Paul Langevin, *Scientia* n° 10, 1911, pp. 31-54.
Max von Laue, *Jahrbucher der Philosophie* 1 (1913, pp. 99-128).
Hermann Weyl, *Raum, Zeit, Materie*, 1918. Note that it is only in the last reference that the *gedankenexperiment* was formulated in terms of twins, a scenario which will meet with an unending success.

[2] Not to mention the presentations, comments and discussions which keep flowing on the internet, let us be content here with a choice of recent articles :
Ø. Grøn, 2013 *Phys. Scr.* **87** 035004
S. Boblest, T. Müller & G. Wünner, 2011 *Eur. J. Phys.* **32** 1117
S. Wortel, S. Malin, M. D. Semon, 2007 *Am. J. Phys.*, **75** 1123

[3] Aesop, "The Hare and the Tortoise", translation Laura Gibbs, 2002, http://mythfolklore.net/aesopica/oxford/237.htm.